\documentclass[aps,preprint]{revtex4}%
\usepackage{amsfonts}
\usepackage{amsmath}
\usepackage{amssymb}
\usepackage{graphicx}
\usepackage{pdfpages}
\usepackage{xcolor}
\usepackage{caption}
\usepackage{lineno,hyperref}
\usepackage{changes}%
\usepackage{subfig}
\providecommand{\U}[1]{\protect\rule{.1in}{.1in}}

\begin{document}
\preprint{HEP/123-qed}
\begin{center}
{\Large Greybody factor and perturbation of a Schwarzschild black hole with string clouds and quintessence}

Ahmad Al-Badawi

Department of Physics, Al-Hussein Bin Talal University, P. O. Box: 20,
71111, Ma'an, Jordan

\bigskip E-mail: ahmadbadawi@ahu.edu.jo

{\Large Abstract}
\end{center}

In this paper, we investigated the Dirac and Klein-Gordan equations, as well
as the greybody factor for a Schwarzschild black hole (SBH) immersed in quintessence and associated with
a cloud of strings. Primarily, we
study the Dirac equation using a null tetrad in the Newman- Penrose (NP)
formalism. Next, we separate the Dirac equation into radial and angular
sets. Using the radial equations, we study the profile of effective
potential by transforming the radial equation of motion into standard Schr%
\"{o}dinger wave equations form through tortoise coordinate. Similarly, we
study the Klein-Gordan equation in this spacetime. The Miller-Good
transformation method is employed to compute the greybody factor of bosons.
To compute the Gerybody factor for fermions, we use the general method of
semi-analytical bounds. We also investigate the effect of string clouds and
quintessence parameters on Hawking radiation. According to the results, the
greybody factor is strongly influenced by the shape of the potential, which
is determined by the model parameters. This is consistent with the ideas of
quantum mechanics; as the potential rises, it becomes harder for the wave to
penetrate.

\section{Introduction}

We are experiencing an accelerating expansion of the universe [1-2], fueled by an unknown exotic energy called dark energy (DE). DE's origin and
essential characteristics remain elusive despite the enormous cosmological
evidence, and this has led to an ongoing debate. Several DE models have been proposed in order to describe the dynamics of the current universe, such as the cosmological constant and the quintessence energy. The quintessence energy [3- 5] is an inhomogeneous, dynamic scalar field which is defined by the equation of state (EoS) $w_{q}=p/\rho $\ with $-1<w_{q}<-1/3$, where $P$ and $\rho $
denote the pressure and energy density, respectively.

In astronomy, quintessence can cause remarkable effects such as the gravitational deflection of distant stars' light [6]. Thus,  we may be able to better understand these effects by studying solutions corresponding to a black hole (BH) surrounded by quintessence, such as Kiselev's [7]. In regards to Kiselev solutions, their rotating counterpart was later constructed [8-9]. The role played by the effects of quintessence on BHs has received considerable attention [10-18].

On the other hand, Letelier [19] proposed a gauge invariant model of a cloud
of strings with the purpose of treating gravity coupled to an array of
strings within the framework of general relativity. Number of studies have
been done about the physics of the cloud of strings [20- 21] and a fluid of
strings [22] within the framework of general relativity. Recently, the null
and timelike geodesics of the Schwarzschild (S) BH with string cloud
background was studied in [23].

In light of recent observations in cosmology, study of the BHs in a
background with quintessence and/or strings as additional sources of gravity
has revealed much about their physics [24-28]. Examining spacetime
characteristics under different kinds of perturbations, including spinor and gauge fields, allows us to understand and analyze its characteristics. In
addition, the greybody factor also plays an important role in calculating
the partial absorption cross section of a BH [29- 32].

Studies have been conducted to compute the greybody factors, to examine the
characteristic bosonic and fermionic quantum radiations, of different BH
backgrounds. Such as the Dirac equation in regular Bardeen BH surrounded by
quintessence [33], in four-dimensional non-Abelian charged Lifshitz black
branes [34], in dRGT massive gravity coupled with nonlinear electrodynamics
[35] and recently in Kerr-like black hole in Bumblebee gravity model [36].

It is the purpose of this paper to study the perturbation and greybody
radiation for a spacetime representing a SBH surrounded by quintessence and
a cloud of strings. Namely, we investigate the Dirac and scalar
perturbations of this spacetime that result in corresponding greybody
radiations. Additionally, we explicitly demonstrate the effect of the cloud
of strings as well as quintessence in this context.

The organization of the paper is as follows. Sect. 2 respectively be devoted to the Dirac and the Klein–Gordon equations in the SBH surrounded by quintessence and a cloud of strings
spacetime. In Sect. 3, we compute the greybody factors of this spacetime for both bosons and fermions. The paper ends with the conclusion in Sect. 4.
\section{Scalar and spinor perturbation}
\subsection{Dirac equation}
Let us start with the spherically symmetric and static BH solution
in the quintessential background, which is surrounded by a cloud of strings
[37-39], namely:
\begin{equation}
ds^{2}=f\left( r\right) dt^{2}-f^{-1}(r)dr^{2}-r^{2}\left( d\theta ^{2}+\sin
^{2}\theta d\phi ^{2}\right)   \label{3}
\end{equation}
where the lapse function $f\left( r\right) $ has the following form
\begin{equation}
f\left( r\right) =1-a-\frac{2M}{r}-\frac{q}{r^{3w_{q}+1}}.  \label{2}
\end{equation}%
in which, $w_{q},$ $M,$ $a,$ and $q$ are the equation of state parameter
(EoS) for quintessence field, mass of the BH, string cloud parameter $(0<a<1)
$, and quintessence parameter respectively. The quintessence parameter is defined as $%
P_{q}=w_{q}\rho _{q}$, with $P_{q}$ and $\rho _{q}$ are the quintessential
energy pressure and density, respectively. The EoS parameter for the quintessence field
has the values $-1<w_{q}<-1/3$ [7]. Here, $w_{q}$ is set to be responsible
for the cosmological acceleration, with $w_{q}=-1$ restoring the
cosmological constant. Metric (\ref{3}) is reduced to SBH if $a$ and $q$ are
not present.

To find massive and massless (fermion) Dirac fields propagating in the space of SBH surrounded by quintessence and a string cloud, the Newman-Penrose formalism will be used [40,41]. The Chandrasekhar-Dirac (CD)
equations [40] in the NP formalism are given by%
\begin{equation*}
\left( D+\epsilon -\rho \right) F_{1}+\left( \overline{\delta }+\pi -\alpha
\right) F_{2}=i\mu _{0}G_{1},
\end{equation*}%
\begin{equation*}
\left( \Delta +\mu -\gamma \right) F_{2}+\left( \delta +\beta -\tau \right)
F_{1}=i\mu _{0}G_{2},
\end{equation*}%
\begin{equation*}
\left( D+\overline{\epsilon }-\overline{\rho }\right) G_{2}-\left( \delta +%
\overline{\pi }-\overline{\alpha }\right) G_{1}=i\mu _{0}F_{2},
\end{equation*}%
\begin{equation}
\left( \Delta +\overline{\mu }-\overline{\gamma }\right) G_{1}-\left( 
\overline{\delta }+\overline{\beta }-\overline{\tau }\right) G_{2}=i\mu
_{0}F_{1},
\end{equation}%
where $F_{1},F_{2},G_{1}$and $G_{2}$ represent the Dirac spinors, $\mu _{0}=%
\sqrt{2}\mu _{p}$ is the mass of the particle $\rho ,\mu ,\epsilon ,\tau
,\gamma ,,\beta ,$ and $\alpha $ are the spin coefficients to be found and
the bar denotes complex conjugation. Now we write the basis vectors of null
tetrad in terms of elements of the metric (\ref{3}) as 
\begin{equation*}
l^{\mu }=\left( \frac{1}{f},1,0,0\right) ,\qquad n^{\mu }=\frac{1}{2}\left(
1,-f,0,0\right) ,
\end{equation*}
\begin{equation}
m^{\mu }=\frac{1}{\sqrt{2}r}(0,0,1,\frac{i}{\sin \theta }),\qquad \overline{m%
}^{\mu }=\frac{1}{\sqrt{2}r}(0,0,1,\frac{-i}{\sin \theta }).\label{7}
\end{equation}%
The directional derivatives in CDEs are defined by $D=l^{\mu }\partial _{\mu
},\Delta =n^{\mu }\partial _{\mu }$ and $\delta =m^{\mu }\partial _{\mu }$.
The spin coefficients can then be computed as 
\begin{align}
\rho & =-\frac{1}{r},\mu =\frac{a-1}{2r}+\frac{M}{r^2}+\frac{q}{2r^{3w_{q}+2}},\epsilon
=\tau =0\qquad   \notag \\
\gamma & =\frac{M}{2r^2}+\frac{q(3w_{q}+1)}{4r^{3w_{q}+2}} ,\beta =-\alpha =\frac{\cot \theta }{2%
\sqrt{2}r}.  \label{8}
\end{align}
Using equations (\ref{7}) and (\ref{8}) CDEs leads to

\begin{equation*}
\left( \emph{D}-\frac{1}{r}\right) F_{1}+\frac{1}{\sqrt{2}r}\mathit{L}%
F_{2}=i\mu _{0}G_{1},
\end{equation*}%
\begin{equation*}
\frac{-f}{2}\left( \mathit{D}^{\dag }-\frac{f^{\prime }}{2f}+\frac{1}{r}%
\right) F_{2}+\frac{1}{\sqrt{2}r}\mathit{L}^{\dag }F_{1}=i\mu _{0}G_{2},
\end{equation*}%
\begin{equation*}
\left( \emph{D}+\frac{1}{r}\right) G_{2}-\frac{1}{\sqrt{2}r}\mathit{L}^{\dag
}G_{1}=i\mu _{0}F_{2},
\end{equation*}%
\begin{equation}
\frac{f}{2}\left( \mathit{D}^{\dag }-\frac{f^{\prime }}{2f}+\frac{1}{r}%
\right) G_{1}+\frac{1}{\sqrt{2}r}\mathit{L}G_{2}=i\mu _{0}F_{1},  \label{D10}
\end{equation}%
where the operators are defined as
\begin{equation*}
\mathit{D}^{\dag }=-\frac{2}{f}\Delta
\end{equation*}%
\begin{equation*}
\mathit{L}=\sqrt{2}r\overline{\delta }+\frac{\cot \theta }{2},
\end{equation*}%
\begin{equation}
\mathit{L}^{\dag }=\sqrt{2}r\delta +\frac{\cot \theta }{2}.
\end{equation}

For the solution of the CDEs (\ref{D10}), we consider the spin- 1/2 wave
function as the form of $F=R\left( r\right) A\left( \theta \right)
e^{i\left( kt+m\phi \right) }$ , where $k$ is the frequency of the incoming
Dirac field and $m$ is the azimuthal quantum number of the wave: 
\begin{equation}
F_{1}=R_{1}\left( r\right) A_{1}\left( \theta \right) e^{i\left( kt+m\phi
\right) },  \notag
\end{equation}%
\begin{equation}
F_{2}=R_{2}\left( r\right) A_{2}\left( \theta \right) e^{i\left( kt+m\phi
\right) },  \notag
\end{equation}%
\begin{equation}
G_{1}=R_{2}\left( r\right) A_{1}\left( \theta \right) e^{i\left( kt+m\phi
\right) },  \notag
\end{equation}%
\begin{equation}
G_{2}=R_{1}\left( r\right) A_{2}\left( \theta \right) e^{i\left( kt+m\phi
\right) }.  \label{10}
\end{equation}%
Substituting Eq. (\ref{10}) into Eqs. (\ref{D10}), the CDEs transform into 
\begin{equation*}
\frac{r}{R_{2}}\left( \emph{D}-\frac{1}{r}\right) R_{1}+\frac{1}{\sqrt{2}%
A_{1}}\mathit{L}A_{2}=ir\mu _{0},
\end{equation*}%
\begin{equation*}
\frac{-rf}{2R_{1}}\left( \mathit{D}^{\dag }-\frac{f^{\prime }}{2f}+\frac{1}{r%
}\right) R_{2}+\frac{1}{\sqrt{2}A_{2}}\mathit{L}^{\dag }A_{1}=ir\mu _{0},
\end{equation*}%
\begin{equation*}
\frac{r}{R_{2}}\left( \emph{D}+\frac{1}{r}\right) R_{1}-\frac{1}{\sqrt{2}%
A_{2}}\mathit{L}^{\dag }A_{1}=ir\mu _{0},
\end{equation*}%
\begin{equation}
\frac{rf}{2R_{1}}\left( \mathit{D}^{\dag }-\frac{f^{\prime }}{2f}+\frac{1}{r}%
\right) R_{2}+\frac{1}{\sqrt{2}A_{1}}\mathit{L}A_{2}=ir\mu _{0}.  \label{D11}
\end{equation}%
To separate equations (\ref{D11}), we define a separation constant. This is
carried out by using the angular equations. In fact, it is already known
from the literature that the separation constant can be expressed in terms
of the spin-weighted spheroidal harmonics. Therefore the radial parts of
CDEs become%
\begin{equation}
\left( \emph{D}+\frac{1}{r}\right) R_{1}=\frac{1}{r}\left( \lambda +ir\mu
_{0}\right) R_{2},  \label{f11}
\end{equation}
\begin{equation}
\frac{f}{2}\left( \mathit{D}^{\dag }+\frac{f^{\prime }}{2f}+\frac{1}{r}%
\right) R_{2}=\frac{1}{r}\left( \lambda -ir\mu _{0}\right) R_{1}.
\label{f12}
\end{equation}
We further assume that \begin{equation}
R_{1}\left( r\right) =\frac{1}{r}P_{1}\left( r\right) ,R_{2}\left( r\right) =%
\frac{\sqrt{f}}{\sqrt{2}r}P_{2}\left( r\right) ,
\end{equation}
then Eqs. (\ref{f11}, \ref{f12}) transform into,%
\begin{equation}
\left( \frac{d}{dr_{\ast }}+ik\right) P_{1}=\sqrt{\frac{f}{r}}\left( \lambda
+ir\mu _{0}\right) P_{2},  \label{f15}
\end{equation}
\begin{equation}
\left( \frac{d}{dr_{\ast }}-ik\right) P_{2}=\sqrt{\frac{f}{r}}\left( \lambda
-ir\mu _{0}\right) P_{1},  \label{f16}
\end{equation}%
where the tortoise coordinate $r_{\ast }$ is defined as $\frac{d}{dr_{\ast }}%
=f\frac{d}{dr}$.

In the end, we combine the solutions in the following way $\psi
_{+}=P_{1}+P_{2},\psi _{-}=P_{2}-P_{1}$ in order to write the equations (\ref%
{f15}, \ref{f16}) in compact form. Hence, we end up with a pair of
one-dimensional Schr\"{o}dinger like equations with effective potentials $%
V_{\pm }$,%
\begin{equation}
\frac{d^{2}\psi _{+}}{dr_{\ast }^{2}}+\left( k^{2}-V_{+}\right) \psi _{+}=0,
\end{equation}%
\begin{equation}
\frac{d^{2}\psi _{-}}{dr_{\ast }^{2}}+\left( k^{2}-V_{-}\right) \psi _{-}=0,
\end{equation}%
\begin{eqnarray}
V_{\pm } &=&\frac{r^{2}L^{3}f}{D^{2}}\pm \frac{rL^{3/2}}{D^{2}}\sqrt{f}\left[
\left( r-M\right) L+3r^{3}\mu _{0}f\right]  \label{p1} \\
&&\mp \frac{r^{3}L^{5/2}}{D^{3}}f^{3/2}\left( 2rL+2r^{3}\mu _{0}+\frac{%
\left( r-M\right) \lambda \mu _{0}}{k}\right)  \notag
\end{eqnarray}
where 
\begin{equation}
L=\left( \lambda ^{2}+\mu _{0}r^{2}\right) ,\qquad D=r^{2}L+\frac{\lambda
\mu _{0}r^{2}}{2k}\left( 1-a-\frac{2M}{r}-\frac{q}{r^{3w_{q}+1}}\right) .
\end{equation}%
The effective potential of the massless Dirac fields (fermions) propagating
in this spacetime can be obtained by setting $%
\mu
_{0}=0$ in (\ref{p1}) namely
\begin{equation}
V_{\pm }=\frac{\lambda ^{2}}{r^{2}}f\pm \frac{\lambda \left( r-M\right) }{%
r^{3}}\sqrt{f}\mp \frac{2\lambda }{r^{2}}f^{3/2}.  \label{p2}
\end{equation}

Expanding the potentials (\ref{p1}) up to order $O\left( \frac{1}{r}\right)
^{3}$ enables us to observe the asymptotic behavior of potentials and the
string cloud parameter as well as the quintessence parameter. The potentials
(\ref{p1}) for SBH with string clouds and quintessence\ (here, we choose $%
w_{q}=-\frac{1}{3}$ ) behave as 
\begin{equation}
V_{\pm }\simeq \left( 1-a\right) \mu _{0}^{2}-\frac{2M\mu _{0}^{2}}{r}\pm
\left( 1-a-q\right) \left( \lambda ^{2}+\left( 1-a-q\right) \frac{\lambda
\mu _{0}}{k}\right) \left( \frac{1}{r}\right) ^{2}+O\left( \frac{1}{r}%
\right) ^{3}.  \label{44}
\end{equation}%
As a consequence of the expanding, the first term is the constant value of
the potential at asymptotic infinity. The second term represents the
monopole-type (or Coulomb-type) potential, while the third term exhibits a
dipole-type potential. The effect of the string cloud parameter as well as
the quintessence parameter can be observed at all orders of $\left( \frac{1}{%
r}\right) $ except the coulomb type term$.$ In the massless case $(%
\mu
=0)$, the potentials (\ref{44}) simplified to%
\begin{equation}
V_{\pm }\simeq \left( 1-a-q\right) \lambda ^{2}\left( \frac{1}{r}\right)
^{2}+O\left( \frac{1}{r}\right) ^{3}.  \label{46}
\end{equation}
\begin{figure}
    \centering
    \subfloat{{\includegraphics[width=6.5cm]{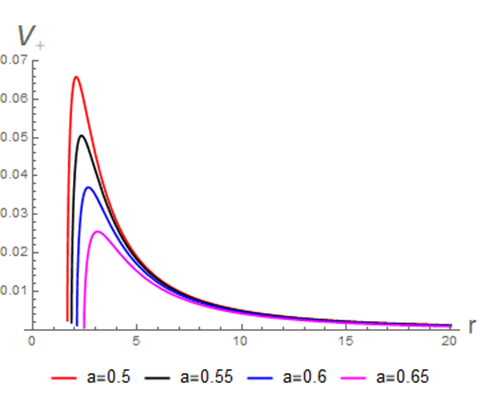} }}%
    \qquad
    \subfloat{{\includegraphics[width=6.5cm]{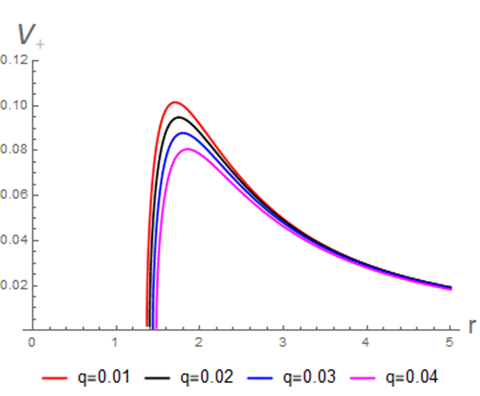} }}%
    \caption{Shows the behaviour of $V_{+}$ (\ref{p2})
for various values of the string cloud parameter $a$  (left plot) and  quintessence parameter $q$ (right plot).  Here, $\lambda = 1, k = 0.2$ and $M = 0.4$.}%
    \label{fig:example}%
\end{figure}
\begin{figure}
    \centering
    \subfloat{{\includegraphics[width=6.5cm]{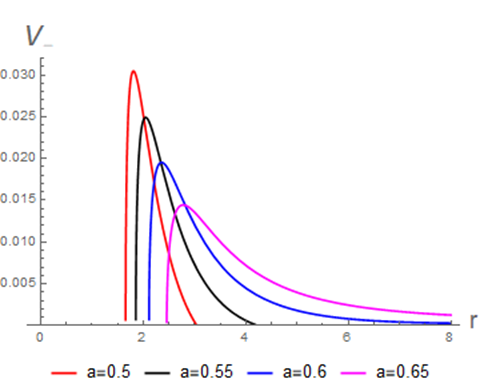} }}%
    \qquad
    \subfloat{{\includegraphics[width=6.5cm]{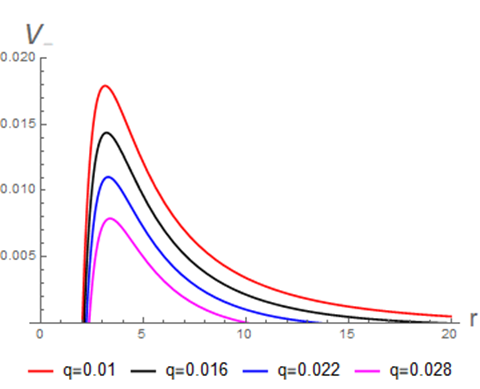} }}%
    \caption{Shows the behaviour of $V_{-}$ (\ref{p2})
for various values of the string cloud parameter $a$  (left plot) and  quintessence parameter $q$ (right plot).  Here, $\lambda = 1, k = 0.2$ and $M = 0.4$.}%
    \label{fig:example}%
\end{figure}
To elaborate the physical behavior of the potentials (\ref{p2}) in the
physical region and to explore the effect of the quintessence parameters $q$
and the string cloud parameter $a$, we make two-dimensional plots of the
potentials for the case $w_{q}=-2/3$ for which $f\left( r\right) =1-a-\frac{%
2M}{r}-qr$. Figures 1 and 2 represent the behavior of the potentials (\ref{p2}) for some specific values of the quintessence and the cloud of strings
parameters. The figures show that the potentials become smaller with higher
values of the quintessence and cloud of strings parameters. The figures
indicate that as the quintessence and string cloud parameters increase, the
potentials decline, so their main effect is to codify potentials and
attenuate their peaks. As a result, we get a hint that when the potentials
peaks diminish, the greybody factor will be higher.
\subsection{Klein-Gordon equation}
The massless scalar field $U(t,r,\theta ,\phi )$ obeys the Klein-Gordon
equation, 
\begin{equation}
\frac{1}{\sqrt{-g}}\partial _{\mu }\sqrt{-g}g^{\mu \nu }\partial _{\nu
}U(t,r,\theta ,\phi )=0,  \label{is1}
\end{equation}%
where $g$ is the determinant of the spacetime metric (\ref{3}), so that $%
\sqrt{-g}=r^{2}\sin \theta $. Here, we are considering a static back ground,
the field equation can be separated as $U=R\left( r\right) Y_{m}^{l}\left(
\theta ,\phi \right) \exp \left( -i\omega t\right) $, where $Y_{m}^{l}$ are
the usual spherical harmonics. Then, the Klein-Gordon equation can be
reduced to a one dimensional Schr\"{o}dinger like equation as follows, 
\begin{equation}
\frac{d^{2}U}{dr_{\ast }^{2}}+\left( \omega ^{2}-V_{eff}\right) U=0,
\label{s1}
\end{equation}%
where $r_{\ast }$ is the tortoise coordinate: $\frac{dr_{\ast }}{dr}=\frac{1%
}{f},$ and $V_{eff}$ is the effective potential given by%
\begin{equation}
V_{eff}=f\left( \frac{\lambda }{r^{2}}+\frac{f^{\prime }}{r}\right) .
\label{v1}
\end{equation}%
where $\lambda =$ $-l(l+1)$, ($l$ is the angular quantum number).

The behaviour of the potential (\ref{v1}) is illustrated in Fig. 3. As one
can see, the potential becomes lower when both quintessence parameter $q$
and string cloud parameters $a$ increase. Once again, this indicates that
the greybody factor will increase as $q$ and $a$ increase.
\begin{figure}
    \centering
    \subfloat{{\includegraphics[width=6.5cm]{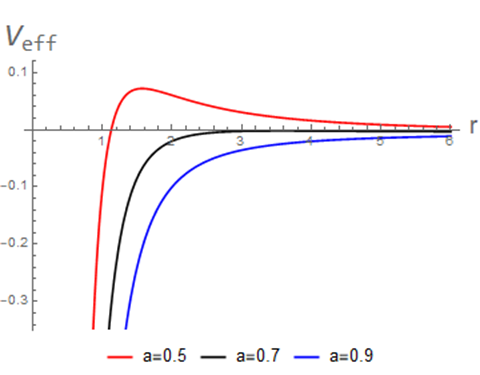} }}%
    \qquad
    \subfloat{{\includegraphics[width=6.5cm]{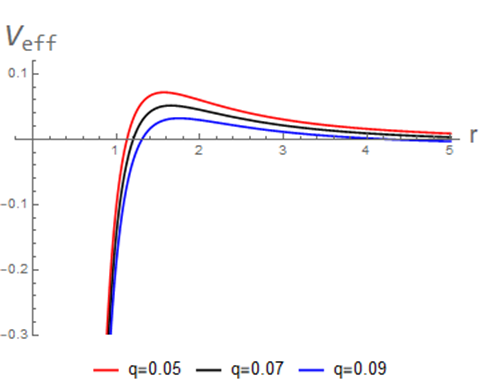}}}%
    \caption{Shows the behaviour of $V_{eff}$ (\ref{v1})
for various values of the string cloud parameter $a$  (left plot) and  quintessence parameter $q$ (right plot).  Here, $\lambda = 1$ and $M = 0.2$.}%
    \label{fig:example}%
\end{figure}

\section{Greybody factor from SBH with string clouds and quintessence}

\subsection{ Greybody factor of bosons}

When quantum effects are considered, BHs can emit thermal radiation, called
Hawking radiation. Greybody factor is one of the quantum quantities of a BH.
It is the fraction of Hawking radiation that can reach spatial infinity. The Miller-Good transformation
method [42-43] will be used to compute the greybody factor of bosons since the
general semi-analytic bounds method yields a measureless greybody factor. Recall that, the Miller-Good transformation method, generates a general bound on quantum transmission probabilities. With this method, a particular transformation is applied to the Schr\"{o}dinger equation (\ref{s1}) to
modify the effective potential (\ref{v1}) and increase the probability that
the Hawking quanta will be transmitted [44].
Hence,  the transmission probability of bosons [14] is given by  
\begin{equation*}
\mathcal{T}\geq \sec h^{2}\left\{ -\frac{1}{2\omega }\int_{r_{h}}^{+\infty
}\left( \frac{f'}{r}+\frac{\lambda }{r^{2}}%
\right) \frac{dr}{f}\right\} 
\end{equation*}
which can be rewritten as%
\begin{equation}
\mathcal{T}\geq \sec h^{2}\left\{ \frac{1}{2\omega }\int_{r_{h}}^{+\infty
}\left( 2M+qr^{-3w_{q}}+\lambda r\right) \left( \frac{1}{r^{4}}\right) \frac{%
dr}{qr^{-3w_{q}-2}+\frac{2M}{r^{2}}-\frac{(1-a)}{r}}\right\} .  \label{in11}
\end{equation}

Here we obtain the gerybody factor for a specific value of EoS parameter
for the quintessence $w_{q}$ $=-2/3$. Therefore, we consider the following
asymptotic expansion 
\begin{equation}
\frac{1}{qr^{-3w_{q}-2}+\frac{2M}{r^{2}}-\frac{(1-a)}{r}}\simeq \frac{1}{q}+%
\frac{1-a}{q^{2}r^{2}}+\frac{\left( 1-a\right) ^{2}-2Mq}{q^{3}r^{2}}+...,
\label{ex1}
\end{equation}%
as a result, we compute the greybody factor (\ref{in11}) as follows%
\begin{multline}
\sigma _{l}\left( \omega \right) \equiv \mathcal{T}\geq \sec h^{2}\left\{ 
\frac{1}{2\omega }\left[ \frac{1}{r_{h}}-\frac{\lambda +a-1}{2qr_{h}^{2}}+%
\frac{1+a^{2}+a(\lambda -2)-\lambda -4Mq}{4q^{2}r_{h}^{3}}\right. \right.
\label{IS45} \\
\left. \left. \frac{2(a-1)Mq-\lambda (1-2a+a^{2}-2Mq)}{4q^{3}r_{h}^{4}}-%
\frac{2M(1-2a+a^{2}-2Mq)}{5q^{3}r_{h}^{5}}\right] \right\} .
\end{multline}
\begin{figure}
    \centering
    \subfloat{{\includegraphics[width=6.5cm]{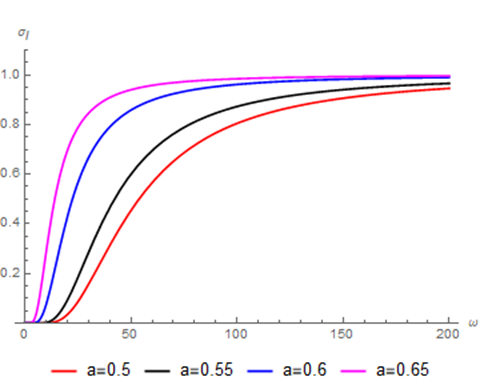} }}%
    \qquad
    \subfloat{{\includegraphics[width=6.5cm]{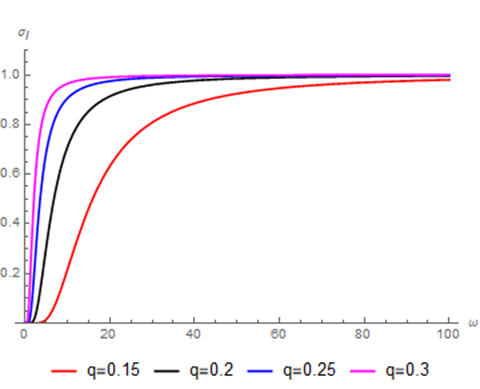}}}%
    \caption{These plots show the bosonic greybody factor (\ref{IS45}) for different values of the string cloud parameter $a$  (left plot) and  quintessence parameter $q$ (right plot).  Here, $\lambda = 1,r = 1$ and $M = 0.4$.}%
    \label{fig:example}%
\end{figure}
The above computed equation (\ref{IS45}) represents the bosonic greybody factor of
SBH surrounded by quintessence and a cloud of strings, which is obtained by
the Miller-Good transformation method. There is no doubt that the specific
form of the greybody factor depends on some parameters that relate to the
potential barrier. The greybody factor bound is actually determined by the
shape of the potential. As in quantum theory, when the potential level
increases, the amplitude of transmission decreases, and as a result the
greybody factor bound decreases. The behavior of greybody factor (\ref{IS45}%
) is depicted in Fig. 4. The plot shows that both the quintessence parameter 
$q$ and the string cloud parameter $a$ are significant for the 
greybody factors. A striking result is that $\sigma _{l}\left( \omega
\right) $ increases with both $q$ and $a$ increasing.
\subsection{Greybody factors of fermions}
Here, we shall derive the fermionic gerybody factor of the neutrinos emitted
from the SBH with string clouds and quintessence. The formula of the general
semi-analytic bounds for greybody factors is \ given by 
\begin{equation}
\sigma _{l}\left( \omega \right) \geq \sec h^{2}\left( \int_{-\infty
}^{+\infty }\wp dr_{\ast }\right) ,  \label{IS37}
\end{equation}

where $\wp$ is the dimensionless greybody factor: 
\begin{equation}
\wp=\frac{\sqrt{\left( h^{\prime}\right) ^{2}+\left(
\omega^{2}-V_{eff}-h^{2}\right) ^{2}}}{2h},  \label{IS38}
\end{equation}

in which $h^{\prime }$ implies the derivation with respect to $r$. By
considering the conditions for $h$ (first, it must be positive and second $%
h\left( +\infty \right) =h\left( -\infty \right) =\omega $), we can simplify
the function as [45]%
\begin{equation}
\wp =\frac{1}{2\omega }V_{eff},  \label{IS39}
\end{equation}

and using the tortoise coordinate as $\frac{dr_{\ast }}{dr}=\frac{1}{f\left(
r\right) }$ then the greybody factor reads 
\begin{equation}
\sigma _{l}\left( \omega \right) \geq \sec h^{2}\left( \frac{1}{2\omega }%
\int_{r_{h}}^{+\infty }V_{eff}\frac{dr}{f}\right) .  \label{gf1}
\end{equation}

To obtain analytical results from (\ref{gf1}) we consider the case of EoS parameter for the quintessence $w_{q}=-1/3$ for which $f\left( r\right) =1-a-%
\frac{2M}{r}-q.$ Indeed, according to recent observational data of Planck
Collaboration [46] the limit of EoS parameter is $-1.16<w_{q}<-0.92$ at the 95 \% confidence level.
This signifies a dark energy model of phantom type with $w_{q}$ less than $-1
$. We are, however, primarily interested in computing the gerybody factor
for a SBH with cloud of strings and quintessence. Thus, we limit our
calculations to $-1\leq w_{q}\leq -1/3.$ We can now substitute the effective
potential (\ref{p2}) into Eq. (\ref{gf1}) to obtain
\begin{equation}
\sigma _{l}^{\pm }\left( \omega \right) \geq \sec h^{2}\frac{\lambda }{%
2\omega }\left( \frac{\lambda }{r_{h}}\pm \frac{1}{\sqrt{1-a-q}}\left( \frac{%
1-2a-2q}{r_{h}}+\frac{M\left( a+q-2\right) }{2\left( 1-a-q\right) r_{h}^{2}}%
\right. \right. \notag
\end{equation}%
\begin{equation}
\left. -\frac{M^{2}}{2\left( 1-a-q\right) ^{2}r_{h}^{3}}-\frac{M^{3}\left(
a+q+4\right) }{8\left( 1-a-q\right) ^{3}r_{h}^{4}}+\frac{M^{4}\left(
5-2a-2q\right) }{8\left( 1-a-q\right) ^{4}r_{h}^{5}}\right),
\end{equation}%
in which $\sigma _{l}^{+}(w)$ and $\sigma _{l}^{-}(w)$ stand for the
greybody factors of the spin-up and spin-down fermions, respectively.
\begin{figure}
    \centering
    \subfloat{{\includegraphics[width=6.5cm]{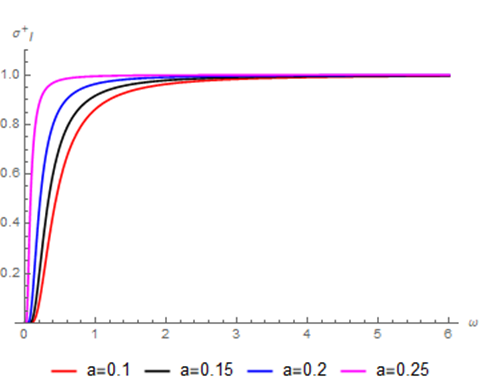} }}%
    \qquad
    \subfloat{{\includegraphics[width=6.5cm]{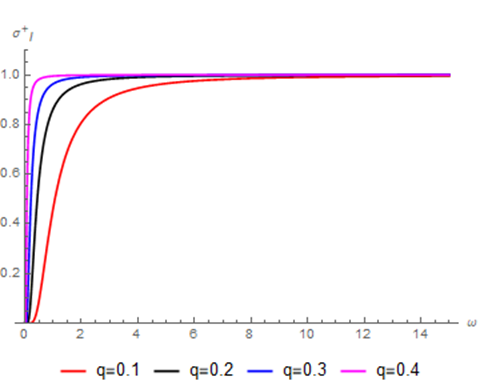}}}%
    \caption{These plots show the fermionic greybody factor spin (+1/2) (32) for different values of the string cloud parameter $a$  (left plot) and  quintessence parameter $q$ (right plot).  Here, $\lambda = 1,r = 1$ and $M = 0.5$.}%
    \label{fig:example}%
\end{figure}
\begin{figure}
    \centering
    \subfloat{{\includegraphics[width=6.5cm]{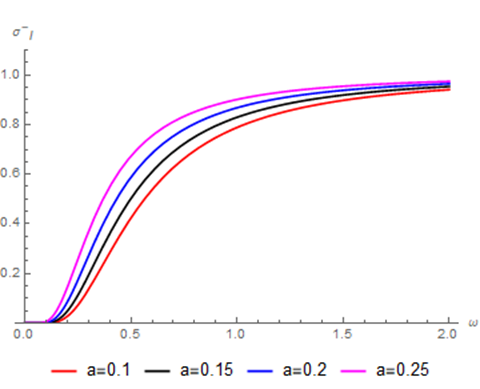} }}%
    \qquad
    \subfloat{{\includegraphics[width=6.5cm]{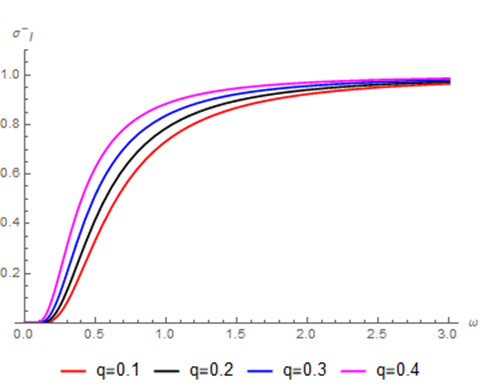}}}%
    \caption{These plots show the fermionic greybody factor spin (-1/2) (32) for different values of the string cloud parameter $a$  (left plot) and  quintessence parameter $q$ (right plot).  Here, $\lambda = 1,r = 1$ and $M = 0.5$.}%
    \label{fig:example}%
\end{figure}
Figures 5 and 6 demonstrate the behavior of spin (+1/2) and spin (-1/2)
under the influence of both the quintessence parameter and string cloud
parameter. According to 5 and 6 the greybody factors increase as $q$ and $a$
increase as with bosons.

\section{Conclusion}

In this paper, we have studied the scalar and spinor perturbations as well
as the greybody radiation in the spacetime of SBH surrounded by quintessence
and a cloud of strings. We examine the Klein-Gordon and Dirac equations,
respectively, for scalar and spinor perturbations. Using the massless
uncharged scalar/spinor field emitted from the SBH with quintessence and a
cloud of strings as Hawking radiation, a Schr\"{o}dinger-like equation with
effective potentials for the radial part of the solution is obtained. Both
quintessence $\left( q\right) $ and string cloud $\left( a\right) $
parameters contribute to the effective potentials. To understand the
physical interpretations and the effect of $q$ and $a$ on the potentials, we
plot the potentials for a variety of parameter values. It is found that the
height of the potentials becomes lower as both $q$ and $a$ increase.
Consequently, we get a clue that the greybody factor bound will rise as the
potential peaks diminish.

In the analysis of the thermal radiation of the SBH surrounded by
quintessence and a cloud of strings, we have utilized the Miller-Good
transformation and the general semi-analytic bounds to obtain the greybody
factors of bosons and fermions, respectively. So, we have shown the effect
of $q$ and $a$ parameters on the Hawking radiation, which can be detected by
observers at spatial infinity. We also supported our results with graphics.
It is found that the greybody factors bound increase as both $q$ and $a$
increase. Thus, the lower the potential, the easier it is for the waves to
be transmitted and therefore, the higher the bound of the greybody factors.
This is in accordance with quantum mechanics.

{\Large Data availability statement:}\newline
My manuscript has no associated data.
\newline

{\Large References}

[1] A.G. Riess et al., Observational evidence from supernovae for an
accelerating universe and a cosmological constant. Astron. J. 116,
1009--1038 (1998)

[2] S. Perlmutter et al., Measurements of omega and lambda from 42 high
redshift supernovae. Astrophys. J. 517, 565--586 (1999)

[3] Steinhardt, P.J., Wang, L., Zlatev, I.: Phys. Rev. D 59, 123504 (1999)

[4] Wang, L., Caldwell, R., Ostriker, J., Steinhardt, P.J.: Astron. J. 530,
17 (2000).

[5] Tsujikawa, S.: Class. Quantum Gravity 30, 214003 (2013).

[6] M. Liu, J. Lu and Y. Gui, Eur. Phys. Jour. C 59 (2009) 107.

[7] V. Kiselev, Class. Quantum Gravity 20 (2003) 1187.

[8] B. Toshmatov, Z. Stuchl\'{\i}k, B. Ahmedov, Eur. Phys. J. Plus 132
(2017).

[9] S.G. Ghosh, Eur. Phys. J. C 76 (2016) 1

[10] S. Fernando, Gen. Relativ. Gravit. 44, 1857 (2012)

[11] K. Ghaderi, Astrophys. Space Sci. 362, 218 (2017).

[12] F. Cicciarella, M. Pieroni, JCAP 1708 (2017).

[13] M. Saleh, B. Thomas, T. Kofane, Eur. Phys. J. C 78:325 (2018).

[14] A. Al-Badawi, S. Kanzi, I. Sakalli, Eur. Phys. J. Plus 135, 219 (2020)

[15] I. AliKhan, A. SultanKhan, S. Islam, Int. J. Mod. Phys. A 35(23),
2050130 (2020).

[16] R. Saadati, F. Shojai, Phys. Rev. D 100, 104041 (2019) 17.

[17] I. Hussaina, S. Alib, Eur. Phys. J. Plus 131, 275 (2016).

[18] K. Ghaderi, B. Malakolkalami, Astrophys. Space Sci. 361 (2016) 161.

[19] P.S. Letelier, Clouds of strings in general relativity. Phys. Rev. D
20, 1294--1302 (1979).

[20] J. Stachel, Phys. Rev. D 21 (1980) 2171.

[21] H. H. Soleng, Gen. Relativ. Gravit. 27 (1995) 367.

[22] L. L. Smalley and J. P. Krisch, Class. Quantum Gravity 14 (1997) 3501.

[23] M. Batool, I. Hussain, Int. J. Mod. Phys. D 26(05), 1741005 (2017).

[24] Efferson de M. Toledo and V. B. Bezerra, Eur. Phys. J. C 78 534 (2018).

[25] Sushant G. Ghosh, Sunil D. Maharaj, Dharmanand Baboolal and Tae-Hun
Lee, Eur. Phys. J. C 78 90 (2018).

[26] Estanislav Herscovich and Martin G. Richarte, Phys. Lett. B 689 (2010)
192.

[27] Tae-Hun Lee, Dharmanand Baboolal and Sushant G. Ghosh, Eur. Phys. J. C
75 (2015) 297

[28] Bouetou Bouetou Thomas, Mahamat Saleh and Timoleon Crepin Kofane, Gen.
Rel. Grav. 44 (2012) 2181.

[29] Y.S. Myung, H.W. Lee, Classical Quantum Gravity 20 (2003) 3533.

[30] T. Harmark, J. Natario, R. Schiappa, Adv. Theor. Math. Phys. 14 (2010)
727.

[31] I. Sakalli, O.A. Aslan, Astropart. Phys. 74 (2016) 73.

[32] H. Gursel, I. Sakalli, Adv. High Energy Phys. 2018 (2018) 8504894.

[33] A. Al-Badawi, I. Sakalli, S. Kanzi, Ann. Phys. 412, 168026 (2020).

[34] H. Gursel, I. Sakalli, Eur. Phys. J. C 80, 234 (2020).

[35] S. Kanzi, S.H.Mazharimousavi, I. Sakalli, Ann. Phys. 422, 168301 (2020).

[36] S. Kanzi, I. Sakall\i , Eur. Phys. J. C 81, 501 (2021).

[37] J.M. Toledoa, V.B. Bezerra, Eur. Phys. J.C 78, 534 (2018). \ \ \ 

[38] J.M. Toledo, V.B. Bezerra, Int. J. Mod. Phys. D 28, 1950023 (2019).

[39] M.M. DiaseCosta, J.M. Toledo, V.B. Bezerra, Int. J. Mod. Phys. D 28,
1950074 (2019)

[40] ] S. Chandrasekhar, The Mathematical Theory of Black Holes, Clarendon,
London, 1983.

[41] E.T. Newman, R. Penrose, J. Math. Phys. 3 (1962) 566.

[42] P. Boonserm, M. Visser, J. Phys. A Math. Theor. 42, 045301 (2009).

[43] S. C. Miller and R. H. Good, Phys. Rev. 91 (1953) 174--179.

[44]  R.M. Wald, General Relativity (The University of Chicago Press, Chicago
and London, 1984).

[45] Y.G. Miao, Z.M. Xu, Phys. Lett. B 772 (2017) 542.

[46] Planck Collaboration:. Astron. Astrophys. 571, A16 (2014).

\end{document}